\documentclass[prl, twocolumn]{revtex4-2}
\usepackage{amssymb}
\usepackage{amsfonts}
\usepackage{amsmath}
\usepackage{bm}
\usepackage{collref}
\usepackage{color}
\usepackage[markup=underlined]{changes}
\usepackage[colorlinks=true, citecolor=blue, urlcolor=blue]{hyperref}
\usepackage{dcolumn}
\usepackage{graphicx}
\usepackage{lmodern}
\usepackage{mathrsfs}
\usepackage{mathtools, slashed}
\usepackage{mathdots}
\usepackage[normalem]{ulem}
\usepackage{physics}
\usepackage{xspace}

\begin{document}
\title{Dynamical effects induced by synthetic SU(1,1) gauge fields}
\author{Chenwei Lv}
\email{lvcw@sustech.edu.cn}
\affiliation{State Key Laboratory of Quantum Functional Materials, Department of Physics, and Guangdong Basic Research Center of Excellence for Quantum Science, Southern University of Science and Technology, Shenzhen, China}
\date{\today}
\begin{abstract}
    Compact gauge fields are powerful tools for controlling quantum dynamics and topological phases of matter. Non-compact gauge fields are particularly relevant for open systems, where particles or energy are exchanged with the environment. Here, we investigate the wavepacket and entanglement dynamics induced by an $\mathrm{SU}(1,1)$ gauge field. The wavepacket undergoes non-Abelian non-compact Bloch oscillations, in which the spectrum of an auxiliary one-dimensional quantum problem determines the transition between stable oscillations and unstable amplification.  In the unstable adiabatic regime, the wavepacket probes an $\mathrm{SU}(1,1)$ topological defect, and its amplification is determined solely by a hyperbolic solid angle in Minkowski space. Threading an $\mathrm{SU}(1,1)$ gauge field through a bosonic tight-binding model yields the bosonic Kitaev model, whose volume or super-volume entanglement scaling is governed by the stability of the gauge field. We further propose an experimental realization of an $\mathrm{SU}(1,1)$ gauge field in a spinor condensate. Our results can be generalized to other non-compact groups and higher dimensions, providing a geometric framework for entanglement control, selective amplification, and engineering energy transfer.
\end{abstract}
\maketitle 

Gauge theory provides a unified framework connecting geometry, electromagnetism, quantum mechanics, and gravity.  
Realizing gauge fields in artificial systems is a long-standing goal of quantum simulation, enabling studies of topological phases and control of strongly correlated many-body systems.
Synthetic U(1) gauge fields have been extensively realized in ultracold atomic systems, allowing for the generation of quantized vortices, the simulation of the Harper-Hofstadter model, and the exploration of both conventional and higher-dimensional quantum Hall physics~\cite{Lin2009, Aidelsburger2013, Mancini2015, Lohse2018, Dalibard2011, Goldman2014}.
Non-Abelian $\mathrm{SU}(2)$ gauge potentials, often implemented through spin-orbit couplings, have also been synthesized in experiments, paving the way for exploring a broad family of transport phenomena and topological phases, including Zitterbewegung, anomalous Hall effects, topological insulators and superconductors~\cite{Lin2011, Zhang2012, Qu2013, Huang2016, Wu2016, Meng2016, Wang2021, Liang2024, Niu2010, Qi2011, Taylor2013}. 

Non-compact groups play a fundamental role in gravity and high energy physics. Gauge groups such as SO(2,2), Poincar\'e, and the rigid affine group arise in exactly solvable models of quantum gravity and in gauge-theoretic formulations of gravity~\cite{Witten1988, Hehl1995, Blagojevi2001}. 
Synthetic non-compact gauge structures offer new routes to quantum control and opportunities to simulate quantum gravity models with accessible many-body platforms. 
Recently, the $\mathbb{R}$ gauge field, also known as the imaginary gauge field, has been synthesized in a spin-orbit coupled Bose-Einstein condensate with spin-dependent loss~\cite{Tao2026}. 
Imaginary magnetic fields and their dynamics have also been proposed and studied~\cite{Ozawa2024, Alminas2026}. 

The $\mathrm{SU}(1,1)$ group is a non-compact counterpart of $\mathrm{SU}(2)$. 
Whereas $\mathrm{SU}(2)$ rotations act on the Bloch sphere, $\mathrm{SU}(1,1)$ transformations generate rotations and boosts on hyperboloids, which can be mapped to a Poincar\'e disk~\cite{Gilmore1990, Novaes2004, Lyu2020}. 
The finite-dimensional realizations of $\mathrm{SU}(1,1)$ are non-unitary.
$\mathrm{SU}(1,1)$ and $\mathrm{SL}(2,\mathbb{C})$ gauge fields in finite-dimensional systems have been studied in the context of non-Hermitian spin-orbit coupling (SOC), and recently in proposals for optical systems~\cite{Zhou2020, Li2022, Ren2022, Kokhanchik2023, Zhao2025, Dong2025, Xiao2026}. 
However, their effects on dynamical phenomena such as Bloch oscillations remain to be fully explored~\cite{Castin1996}.
Meanwhile, $\mathrm{SU}(1,1)$ also governs unitary dynamics in infinite-dimensional Hilbert spaces, including inverted harmonic oscillators, Bogoliubov excitations, and scale-invariant quantum gases~\cite{Subramanyan2021, Lyu2020, Werner2006, Lv2020}. 
These realizations motivate extending the gauge-theoretic perspective to bosonic many-body systems, providing a unified picture of particle creation, dynamical instabilities, and entanglement phase transitions.

A Hamiltonian with an $\mathrm{SU}(1,1)$ spin coupled to the motional degree of freedom in a translation-invariant one-dimensional system is written as
\begin{equation}
    \hat H(p) = \epsilon(p){\mathbb{I}}_2 + \xi_0(p)\hat K_0 + \xi_1(p)\hat K_1 + \xi_2(p)\hat K_2.~\label{eq:soc}
\end{equation}
We set $\hbar$ and the energy unit to unity.
$\hat K_{0,1,2}$ are the generators of the $\mathfrak{su}(1,1)$ algebra, satisfying the commutation relations
\begin{equation}
    [\hat K_0,\hat K_1]=i\hat K_2,\, [\hat K_1,\hat K_2]=-i\hat K_0,\, [\hat K_2,\hat K_0]=i\hat K_1.~\label{eq:SU11}
\end{equation} 
In a finite-dimensional representation, $\hat K_0=\hat \sigma_x/2$, $\hat K_1=i\hat \sigma_y/2$ and $\hat K_2=i\hat \sigma_z/2$. 
Therefore, $\hat H(p)$ could be regarded as a non-Hermitian SOC Hamiltonian.
The spectrum of $\hat H$ is $\epsilon_\pm = \epsilon \pm \xi/2$, where $\xi=\sqrt{\xi_0^2-\xi_1^2-\xi_2^2}$ is the $\mathrm{SU}(1,1)$ field strength. 
In infinite-dimensional representations, $\hat K_\mu$'s can be chosen Hermitian.

\begin{figure}[h]
    \includegraphics[width=1\linewidth]{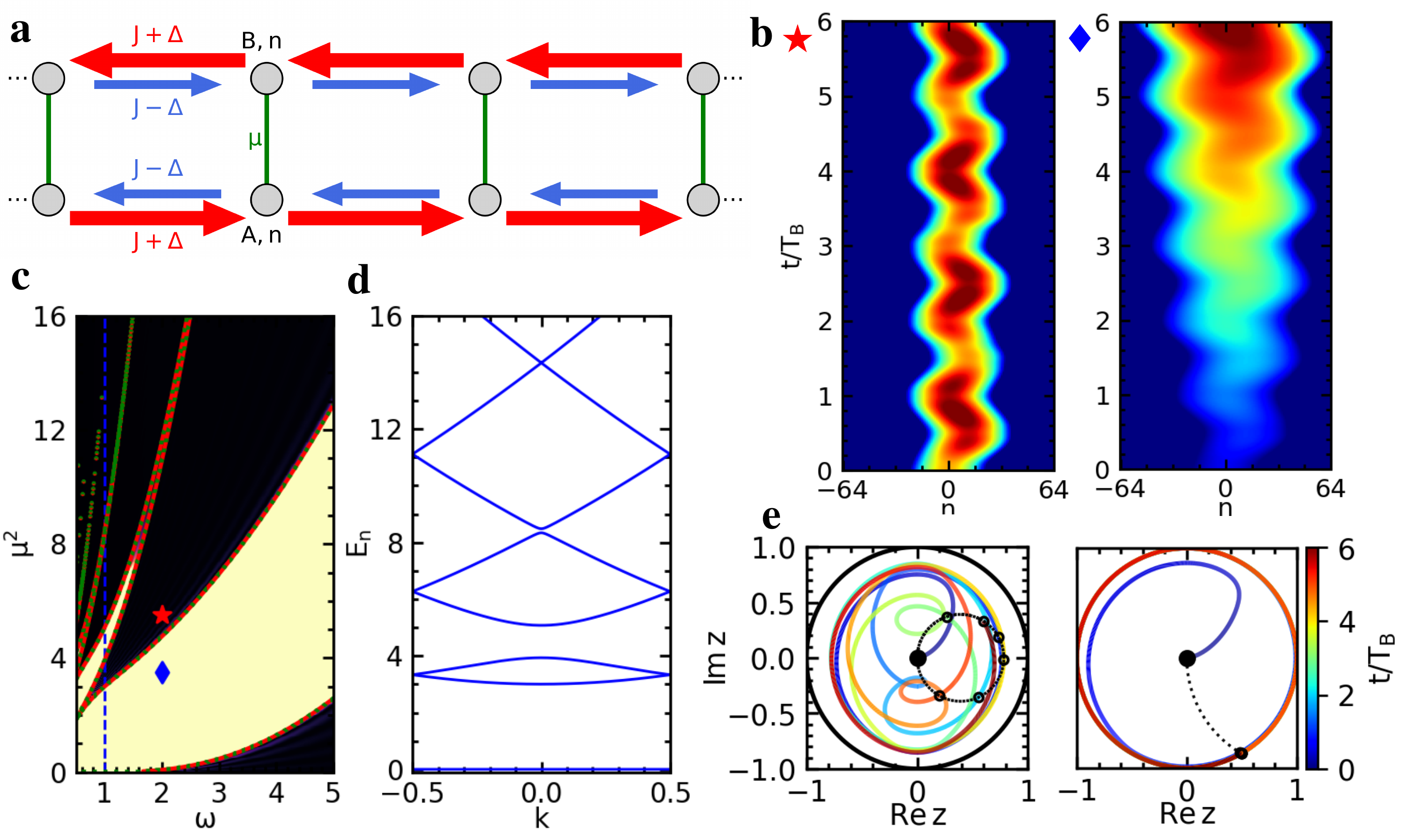}
    \caption{
        (a) Schematic plot of the coupled HN chains. 
        (b) Stable (unstable) wavepacket dynamics for $\omega=2$, $\mu^2=5.5(3.5)$ marked by the red star (blue diamond) in (c). 
        (c) Density plot of the total wavepacket norm at $t=6T_B$ with $\Delta=1,J=6$ and different $\omega$ and $\mu^2$. 
        Red solid and green dotted curves are the phase boundaries obtained from Eq.~\eqref{eq:order} and Eq.~\eqref{eq:Whittaker}, respectively. 
        The blue dashed line denotes the parameters in (d). 
        (d) Energy band of Eq.~\eqref{eq:Whittaker} with $\Delta=\omega=1$. 
        (e) Internal spin dynamics on the Poincar\'e disk. Time is indicated by the color scale and black dots (dotted curves) represent the states at $t/T_B\in\mathbb{Z}$ (effective dynamics).}
    \label{fig:Bloch}
\end{figure}

We consider coupled Hatano-Nelson (HN) chains with opposite chirality, as shown in Fig.~\ref{fig:Bloch}(a). 
The equation of motion reads
\begin{equation}
    i\partial_t\begin{pmatrix}
        \psi_{A,n} \\ \psi_{B,n}
    \end{pmatrix}
    =\begin{pmatrix}
        t_+\psi_{A,n-1} + t_-\psi_{A,n+1}+\mu \psi_{B,n}\\ 
        t_-\psi_{B,n-1} + t_+\psi_{B,n+1}+\mu \psi_{A,n}
    \end{pmatrix},~\label{eq:cHN_real}
\end{equation}
where $t_\pm = J\pm\Delta$, and $\mu$ is the interchain coupling.
The HN chain is a well-known non-Hermitian model with asymmetric hopping that exhibits the non-Hermitian skin effect and chiral transport~\cite{Hatano1996, Wang2018, Weidemann2020}.
Its continuum limit also admits a description in terms of dynamics in curved space~\cite{Lv2022}.
Eq.~\eqref{eq:cHN_real} exhibits the critical skin effect, an exponential spectral sensitivity to small $\mu$ as the system size increases~\cite{Li2020}. 
Under periodic boundary conditions, the momentum space Hamiltonian becomes 
\begin{equation}
    \hat H_{\rm c}(p) = 2J\cos(p)\mathbb{I}_2 + 2\Delta\sin(p)i\hat \sigma_z + \mu \hat \sigma_x.~\label{eq:cHN}
\end{equation}
Eq.~\eqref{eq:cHN} is identified with Eq.~\eqref{eq:soc} when $\epsilon=2J\cos(p)$, $(\xi_0,\xi_1,\xi_2)=(2\mu,0,+4\Delta\sin p)$. 
The right eigenstates are $\ket{\pm}=(2i\Delta\sin(p)\pm\xi/2,\mu)^T$.
A wavepacket evolves as $\psi(n,t)=\sum_{p,s=\pm}c_{s}(p,0)\ket{s}e^{-ipn-i\epsilon_{s}(p)t}$, where $c_{\pm}$ are the momentum space wavefunctions obtained from the corresponding biorthogonal left eigenstates~\cite{Brody2013}.
For wavepackets narrowly distributed in momentum space around $p_0$, the evolution of its internal spin is approximately governed by $\hat H_{\rm c}(p_0)$, and is stable (unstable) if $\xi^2(p_0)>(<)0$.
Derivatives of $\hat H_{\rm c}(p)$ are relevant when the position and shape of the wavepacket are of interest.

A linear potential moves the wavepacket in the momentum space.
Applying $V_n=n\omega$ to both chains in Eq.~\eqref{eq:cHN_real} changes $p$ linearly, $\partial_t p=\omega$.
In the absence of an SU(1,1) gauge field, $\Delta=\mu=0$, the internal spin remains stationary. 
The group velocity $-\partial_p\epsilon$ changes periodically, which produces ordinary Bloch oscillations with period $T_B=2\pi/|\omega|$ and amplitude $2|J/\omega|$~\cite{Castin1996, Hartmann2004, Zhang2025}.
When $\Delta\neq 0$ and $\mu=0$, the Bloch oscillation becomes non-Hermitian and non-reciprocal, which admits a description in terms of complex trajectories~\cite{Longhi2009, Longhi2015}. 
For non-zero $\mu$ and $\Delta$, the internal spin evolves with $\mathrm{SU}(1,1)$ dynamical symmetry. 
The internal Hamiltonians $\hat H_{\rm c}(p)$ encountered at different momenta generally do not commute, resulting in non-Abelian Bloch oscillations with dynamical instabilities. 
Fig.~\ref{fig:Bloch}(b,c) shows the evolution of the wavepacket and its norm after $6T_B$ at various parameters, with an initially normalized Gaussian wavepacket $\psi_0(n)\propto e^{-n^2/4\sigma^2}(1,1)^T$, $p_0=0$.
Depending on $\mu$, $\Delta$ and $\omega$, the one-cycle evolution operator can support either stable oscillation or exponential amplification (Fig.~\ref{fig:Bloch}(b)), separated by phase boundaries shown in Fig.~\ref{fig:Bloch}(c).
Fig.~\ref{fig:Bloch}(e) maps the internal spin dynamics to the Poincar\'e disk~\cite{Zhang2022}. 
Black markers denote the states at stroboscopic times, and dashed curves show evolution generated by the Floquet Hamiltonian $\hat H_F$ defined below.
Although $\hat H_c(p(t))$ has period $T_B$, the internal spin generally evolves with a different period set by $\hat H_F$, or becomes aperiodic if $\hat H_F$ is unstable.

The stability of Bloch oscillations is determined by the $\mathrm{SU}(1,1)$ Floquet evolution operator 
\begin{equation}
    \hat{\mathcal{U}}=\mathcal{T}{\rm exp}(-i\int_{t=0}^{T_B}H_{\rm c}(\omega t)dt), \label{eq:order}
\end{equation}
of the internal spin.
$\mathcal{T}$ denotes time ordering. 
$\hat H_F = i\log\hat{\mathcal{U}}/T_B$ defines the Floquet Hamiltonian governing the effective dynamics at stroboscopic times~\cite{Dalibard2014}.
Dynamics is exponentially unstable if $|{\rm Tr}(\hat{\mathcal{U}})| > 2$. 
This condition is related to energy spectrum of a periodic potential. 
We define $\tilde\psi=e^{+2iJ\sin(\omega t)/\omega}\psi$.
Applying $i\partial_t$ to the Schr\"odinger equation with Hamiltonian Eq.~\eqref{eq:cHN} yields
\begin{equation}
   -\partial_t^2\tilde\psi+[4\Delta^2\sin^2(\omega t)+2\Delta\omega\cos(\omega t)\sigma_z]\tilde\psi = \mu^2\tilde\psi. ~\label{eq:Whittaker}
\end{equation}
Treating $t$ as a spatial coordinate, this is a time-independent Schr\"odinger equation with potential $V_\pm=4\Delta^2\sin^2(\omega t)\pm 2\Delta\omega\cos(\omega t)$ and energy $\mu^2$, known as the Whittaker-Hill equation~\cite{Hemery2010}. 
$V_\pm$ have the same spectrum, as they are related by a translation of $\pi/\omega$.
Whether $\mu^2$ falls inside a gap determines the behavior of $\tilde\psi$ at $|t|\to\infty$. 
Therefore, the spectrum of the Whittaker-Hill equation determines the stability of the Bloch oscillations in the coupled HN chains, band interiors (gaps) give bounded evolution (exponential instability).
In Fig.~\ref{fig:Bloch}(d), we plot the energy band of Eq.~\eqref{eq:Whittaker} at $\omega=\Delta = 1$.
The band interiors (gaps) are inside the stable (unstable) phases, as shown along the blue dashed line in Fig.~\ref{fig:Bloch}(c).
We also plot the curves representing $|{\rm Tr}(\hat{\mathcal{U}})|=2$ (red solid), and the energy band edges of Eq.~\eqref{eq:Whittaker} (green dotted). 
They agree with the phase boundaries. 
This approach extends to a wide range of Hamiltonians with $\mathrm{SU}(1,1)$ dynamical symmetry (supplementary materials).  

\begin{figure}[t]
    \includegraphics[width=1\linewidth]{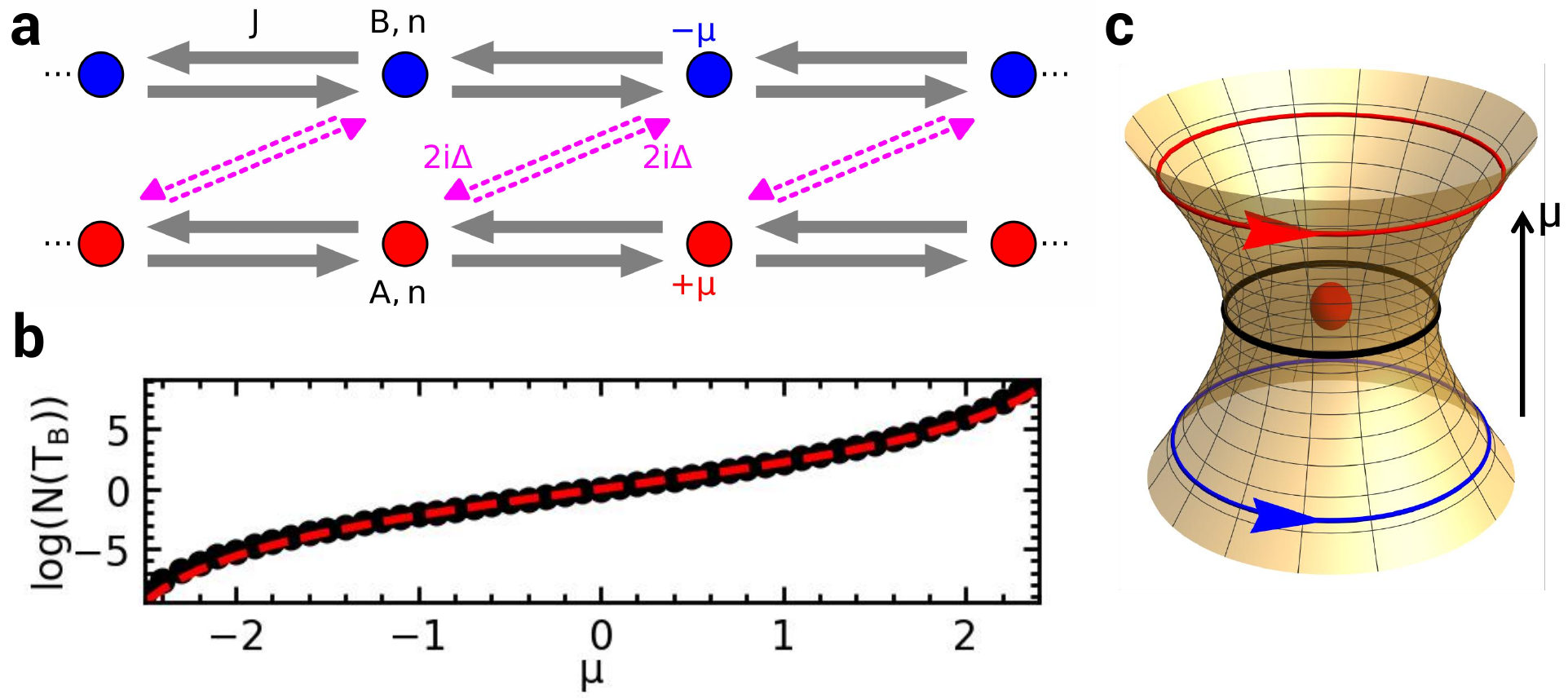}
    \caption{
        (a) Schematic plot of the lattice model considered in Eq.~\eqref{eq:cHN2}. 
        (b) Norm of the wavepacket at $t=T_B$ on a log scale, $\Delta=1.5$, $J=0.1$. 
            Red dashed curve denotes the geometric factor $\exp(2\pi\mu/\Gamma)$ obtained from the hyperbolic solid angle. 
        (c) Schematic plot of the topological defect and the adiabatic trajectories of the SU(1,1) spin. The black loop at $\eta=0$ is a reference with zero imaginary Berry phase.}
    \label{fig:Adiabatic}
\end{figure} 

A particularly interesting limit is when the gradient of the linear potential is much smaller than the energy gap $|\omega|\ll |\xi(p)|$, such that the internal spin follows the adiabatic path as the wavepacket sweeps across the Brillouin zone.
Similar to $\mathrm{SU}(2)$, a topological defect exists and generates geometric effects when a finite area is enclosed by a cyclic path in the parameter space~\cite{Garrison1988, Sun1993, Gadway2023}. 
To this end, we introduce an additional coupling term,
\begin{equation}
    \hat H'_{\rm c}(p) = \hat H_{\rm c}(p) + 2\Delta\cos(p)i\hat \sigma_y -i\Gamma\mathbb{I}_2,~\label{eq:cHN2}
\end{equation}
and consider the unstable regime $|\mu|<2|\Delta|$.
Eq.~\eqref{eq:cHN2} after a unitary transformation $\hat\sigma_{x,y,z}\to\hat\sigma_{z,x,y}$ is represented in Fig.~\ref{fig:Adiabatic}(a).
We also introduce a global decay $\Gamma=\sqrt{4\Delta^2-\mu^2}$, such that ${\rm Im}(\epsilon_\pm)\leq 0$ for all $p$.
In Fig.~\ref{fig:Adiabatic}(b), we calculate the norm of $\psi_0$ at $t=T_B$ for $\omega=+0.01$ and different $\mu$, $N(T_B)=\sum_n|\psi(n,T_B)|^2$. 
Exponential growth is observed for $\mu>0$, even though the spectrum $\epsilon_\pm(p)$ of $\hat H'_{\rm c}(p)$ is purely dissipative. 
As represented by the red dashed curve, the one-cycle growth/decay $\log(N(T_B)/N(0))$ matches the hyperbolic solid angle, $\Omega_M=2\pi\mu/\Gamma$, of the region enclosed by the adiabatic trajectory at different $\mu$'s and the reference loop as plotted in Fig.~\ref{fig:Adiabatic}(c).
The calculation of $\Omega_M$ from embedding is shown in the supplementary materials.

To show the defect, we regard $\xi_{0,1,2}$ in Eq.~\eqref{eq:soc} as parameters, and compute the Berry phase in the parameterization $(\xi_0,\xi_1,\xi_2) = \bar\xi (\sinh\eta,\cosh\eta\cos\varphi,\cosh\eta\sin\varphi)$, where $\bar\xi=\sqrt{\xi_1^2+\xi_2^2-\xi_0^2}>0$.
The right eigenvectors are $\ket{\pm}_R=(\alpha_\pm,\pm\alpha_\pm^*)_R^T$, where $\alpha_\pm=e^{-i\varphi/2}(\cosh(\eta/2)\mp i\sinh(\eta/2))/\sqrt{2}$. 
$\ket{\pm}_L(\eta)=\ket{\pm}_R(-\eta)$ are the left eigenvectors.
The standard biorthogonal approach for Berry connections of non-Hermitian systems considers the upper branch and shows $A_\mu=i\langle +|_L\partial_\mu |+\rangle_R$~\cite{Garrison1988, Sun1993, Brody2013}, 
\begin{equation}
    A_{\bar\xi}=A_\eta = 0,\quad A_\varphi = -i\sinh(\eta)/2.
\end{equation}
An imaginary Berry connection generates an imaginary Berry phase, which produces geometrical amplification/attenuation.
We can transform the connection one-form $A_\varphi d\varphi$ back to the $\xi_i$-parameter space and take its curl, obtaining
\begin{equation}
    \vec B = \frac{i}{2\bar\xi^3}(\xi_0,\xi_1,\xi_2)^T.
\end{equation}
This magnetic field can be regarded as generated by a monopole with charge $i/2$ and magnetic potential $1/\bar\xi$. 
Although $|\vec B|$ diverges at the exceptional cone at $\xi=0$, the monopole locates at the origin. 
In the stable regime, $\vec B$ can be identified with an electromagnetic field generated by a topological instanton in a $2+1$ dimensional Minkowski spacetime~\cite{Zhang2018, Ran2025}. 
Its analytical continuation outside of the light cone provides the geometrical amplification we observed and can be directly probed through wavepacket dynamics.

We now provide infinite-dimensional, Hermitian realizations of the $\mathrm{SU}(1,1)$ gauge field in bosonic systems. 
We use the bosonic Kitaev model (BKM) as an example.  
For a single bosonic mode $\hat a$, $[\hat a,\hat a^\dag]=1$, define
\begin{equation}
    \hat{\mathcal{K}}_0=(\hat a^\dag \hat a+\hat a\hat a^\dag)/4,\, \hat{\mathcal{K}}_{+}=\hat a^{\dag 2}/2, \hat{\mathcal{K}}_\pm=\hat{\mathcal{K}}_{1}\pm i\hat{\mathcal{K}}_{2},
\end{equation}
$\hat{\mathcal{K}}_{-}=\hat{\mathcal{K}}_{+}^\dag$.
The $\mathfrak{su}(1,1)$ algebra in Eq.~\eqref{eq:SU11} is satisfied. 
Such symmetry allows us to thread an $\mathrm{SU}(1,1)$ gauge field through a quantum system with local bosonic degrees of freedom using local squeezing transformations.
The BKM Hamiltonian on $L$ sites is written as~\cite{McDonald2018,Slim2024,Busnaina2024,Lv2024}
\begin{equation}
    \hat H_b=\sum\nolimits_n e^{i\theta}J\hat T_n+ e^{i\phi}\Delta \hat S_n+h.c.,\label{eq:bKM}
\end{equation}
where $\hat T_n = \hat a_n^\dag \hat a_{n+1}$, and $\hat S_n = \hat a_n^\dag \hat a_{n+1}^\dag$ are the nearest neighbor tunnelings and pair creation terms, respectively. 
The equation of motion for the Nambu-Bogoliubov spinor $\hat \psi_n=(\hat a_n,\hat a_n^\dag)^T$ is 
\begin{equation}
    i\partial_t\hat\psi_n=\hat M_+\hat\psi_{n+1}+\hat M_-\hat\psi_{n-1},~\label{eq:bKM2}
\end{equation}
where $\hat M_\pm=\pm iJ\sin(\theta){\mathbb{I}_2}+\hat M$, $\hat M = 2(\cos(\theta)J\hat K_0+\Delta(\sin(\phi)\hat K_1+\cos(\phi)\hat K_2))$. Here, $\hat K_0=\hat \sigma_z/2$, $\hat K_1=i\hat \sigma_x/2$ and $\hat K_2=i\hat \sigma_y/2$. 
When $-\theta=\phi=\pi/2$, it becomes the bosonic Kitaev chain~\cite{McDonald2018,Slim2024,Busnaina2024}. $\hat M_\pm$ on the link can define an $\mathrm{SU}(1,1)$ group element, and the gauge transformations are local Bogoliubov transformations that redefine the local vacuum.

\begin{figure}[t]
    \includegraphics[width=1\linewidth]{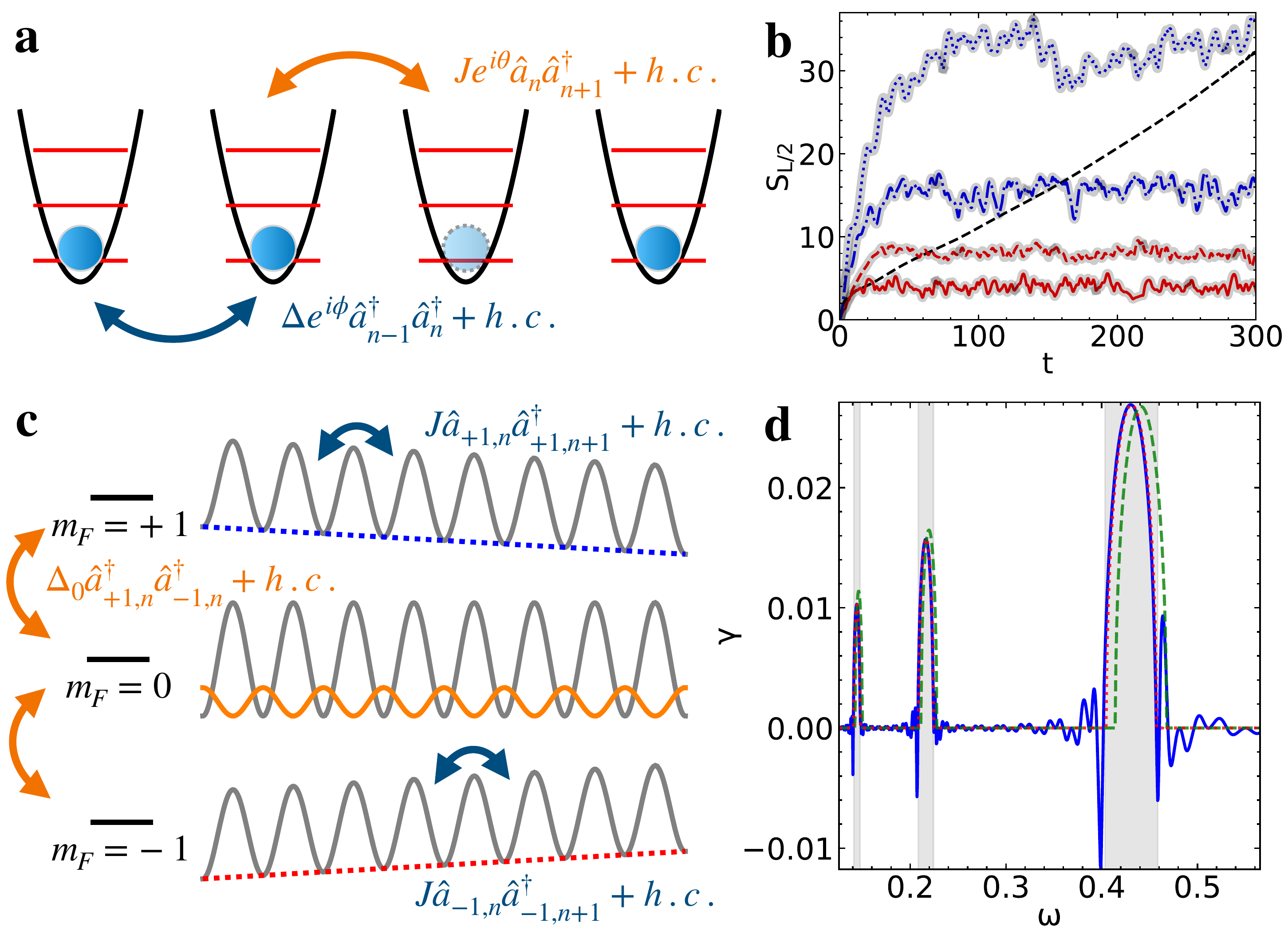}
    \caption{
        (a) Schematic plot of the BKM. 
        (b) Dynamics of half-chain entanglement entropy in the BKM with a vacuum initial state. $\Delta=0.5$, $J=1.0$, $\phi=-0.35$. From top to bottom: $\theta=1.2,L=20$ (dotted), $\theta=1.2,L=14$ (dash-dotted), $\theta=1.0,L=20$ (dashed), $\theta=1.0,L=10$ (solid). Black dashed curve has $J=\phi=0$, $\Delta=0.01$, and $L=20$. Gray thick curves are obtained from locally squeezed initial states in the tight-binding model. 
        (c) Schematic plot of a spin-1 Bose gas in tilted optical lattices.  
        (d) Particle growth rate in $m_F=\pm 1$ obtained from fits to the GP simulation results (blue solid curve), the effective SU(1,1) spin precession of Eq.~\eqref{eq:ladder} (green dashed curve), and after including higher Wannier bands and longer-range terms (red dotted curves). The gray region denotes exponential fits with $R^2\geq 0.95$.
        The parameters are chosen as $q=0.04$, $c_0=0.2$, $c_2=0.03$, $V_0=5$. 
        GP simulation is performed on $36$ unit cells for $24T_B$ and the GP wavefunction has unit mean density. }
    \label{fig:bosonic}
\end{figure}

The $\mathrm{SU}(1,1)$ gauge field in Eq.~\eqref{eq:bKM} is trivial and can be gauged away on a finite open chain. 
When $\Delta < J$, we define $\hat G_\pm=\mp i\hat M_\pm/J_{\rm eff}$, $J_{\rm eff} = \sqrt{J^2-\Delta^2}$.
Since $\hat G^\dag_\pm\hat \sigma_z \hat G_\pm=\hat \sigma_z$, and ${\rm det}(\hat G_\pm)=1$, $\hat G_\pm$ are $\mathrm{SU}(1,1)$ group elements. 
$\hat G_-\hat G_+=\mathbb{I}_2$.
Define a site-dependent $\mathrm{SU}(1,1)$ transformation $\hat{\mathcal{G}}_n$ for each $\hat \psi_n$, such that
\begin{equation}
    \hat{\mathcal{G}}_n^\dag\hat\psi_n\hat{\mathcal{G}}_n=\hat G^n_-\hat\psi_n.
\end{equation}
After the transformation, Eq.~\eqref{eq:bKM2} reduces to $i\partial_t\hat\psi_n=iJ_{\rm eff}(\hat\psi_{n+1}-\hat\psi_{n-1})$. 
Therefore, Eq.~\eqref{eq:bKM} is unitarily equivalent to a bosonic tight-binding model with tunneling strength $J_{\rm eff}$, $\theta_{\rm eff} =\pi/2$ and $\Delta_{\rm eff}=0$.
When $|{\rm Tr}(\hat G_-)|=2J|\sin(\theta)|/J_{\rm eff}<(>)2$, $\hat{\mathcal{G}}_n$ is stable (unstable) as $|n|$ increases.
When $\Delta > J$, Eq.~\eqref{eq:bKM} is unitarily connected to a squeezing model with $\Delta_{\rm eff} = \sqrt{\Delta^2-J^2}$, $\phi_{\rm eff}=J_{\rm eff}=0$ (supplementary materials).

This $\mathrm{SU}(1,1)$ gauge structure is particularly useful for understanding entanglement dynamics in the BKM. 
Since Eq.~\eqref{eq:bKM} can be transformed into a tight-binding or squeezing model through local unitary transformations, the entanglement dynamics of a vacuum input state in the original model is identical to that of a locally squeezed initial state in the transformed model. 
For $\Delta>J$, the mapped squeezing model can generate unbounded linear entanglement growth at long times~\cite{Bianchi2018}.
For $J>\Delta$, passive dynamics generated by the tight-binding model redistributes the initial squeezing into nonlocal supermodes and the entanglement growth saturates.
For one squeezed supermode $\hat b=\sum b_n\hat a_n$, $\sum_n|b_n|^2=1$, with a squeeze parameter $r$, the half-chain entanglement entropy is (supplementary materials)
\begin{equation}
    S_{L/2}=(n_{\rm th}+1)\ln(n_{\rm th}+1)-n_{\rm th}\ln(n_{\rm th}),
\end{equation}
where $2n_{\rm th}=\sqrt{(\eta e^{-2r}+1-\eta)(\eta e^{+2r}+1-\eta)}-1$. $\eta=\sum_{n<L/2}|b_n|^2$ is the mode weight on one half of the system.
When $\eta=1/2$, $S_{L/2}\approx r$ as $r\to\infty$.
For multiple squeezed supermodes, individual contributions cannot generally be added, because their restrictions to a subsystem are not orthogonal and can interfere. 
The exact $S_{L/2}$ follows from the symplectic eigenvalues of the reduced covariance matrix~\cite{Marcos2018}.
Nevertheless, we can qualitatively estimate the total $S_{L/2}$ as a sum of the individual contributions.
When $\hat G_\pm$ is stable, the total squeezing resource of the locally squeezed initial state grows at most linearly with $L$.
When $\hat G_\pm$ is unstable, the initial state becomes more squeezed as $L$ increases. 
The total squeezing resource scales as $L^2$, which indicates the same scaling of the entanglement entropy at equilibrium.
This effect is known as the super-volume law scaling in the study of BKC~\cite{Lee2024}. 
Here, we show that these entanglement phases is originated from an interplay between a non-compact gauge field and passive dynamics. 
Such a scheme can be directly generalized to any other systems with local $\mathrm{SU}(1,1)$ or other non-compact symmetry groups, such as a bosonic ladder, and to higher dimensions. 

Our second example is a bosonic ladder with squeezing terms. Its Hamiltonian is written as 
\begin{equation}
    \hat H_{\rm l} = \sum\nolimits_{n,s=\pm} \epsilon_0\hat N_{s,n}+\big[J\hat T_{s,n}+\Delta_0 \hat S_{n}'+h.c.\big],~\label{eq:ladder}
\end{equation}
where $\hat N_{\pm,n}=\hat a^\dag_{\pm,n} \hat a_{\pm,n}$, $\hat T_{\pm,n}=\hat a^\dag_{\pm,n} \hat a_{\pm,n+1}$,
and $\hat  S_{n}'=\hat a^\dag_{+,n} \hat a^\dag_{-,n}$. 
$\hat T_{\pm,n}$ and $\hat S'_{n}$ are intrachain tunneling of particles/holes and the pair creation of particles in both chains. $\hat S'_{n}$ can also be understood as interchain tunnelings with particles converted into holes.
The equation of motion of the Nambu-Bogoliubov spinor $\hat\Psi_n=(\hat a_{+,n},\hat a_{-,n}^\dag)^T$ in the momentum space has the form of Eq.~\eqref{eq:soc}, with $(\xi_0,\xi_1,\xi_2) = (2\epsilon_0+4J\cos(p),0,2\Delta_0)$.
A linearly varying potential with opposite signs, $\hat H_E = \omega\sum_n n (\hat N_{+,n}-\hat N_{-,n})$ 
applies an identical electric field to particles and holes in $\hat\Psi_n$, therefore delivers non-Abelian non-compact Bloch oscillations in operator dynamics. 

The $\mathrm{SU}(1,1)$ gauge field and the bosonic ladder have a natural realization in a spinor condensate~\cite{Ho1998, Kawaguchi2012}. 
The Hamiltonian of a one-dimensional spin-1 Bose gas can be written as 
\begin{equation}
  \begin{split}
    &\hat H=\int dx \bigg[|\partial_x\hat \psi_a|^2+V|\hat \psi_a|^2+\frac{c_0}{2}\hat \psi_a^\dag \hat \psi_{a'}^\dag \hat \psi_a\hat \psi_{a'} \\&+\frac{c_2}{2}\hat \psi_a^\dag\hat \psi_{a'}^\dag \vec S_{ab}\cdot \vec S_{a'b'}\hat \psi_{b}\hat \psi_{b'}+q(|\hat \psi_1|^2+|\hat \psi_{-1}|^2)\bigg].
  \end{split}~\label{eq:spinor}
\end{equation}
The repeated indices are summed over $m_F=\pm 1,0$. 
$c_2$ ($c_0$) is the spin-(in)dependent interaction strength, 
$\vec S$ denotes the spin-1 matrices and $q$ is the strength of quadratic Zeeman shift. 
When $c_2>0$ and $q>0$, the mean-field ground state of Eq.~\eqref{eq:spinor} is polar and condensed in $m_F=0$~\cite{Ho1998}. 
$\hat\psi_0(x)$ can be replaced by a c-number, $\Phi_0(x)e^{-i\bar\mu t}$, where $\bar\mu$ is the chemical potential of the Gross-Pitaevskii (GP) ground state. The spin-dependent interaction induces an $\mathrm{SU}(1,1)$ gauge field within the $m_F=\pm 1$ manifold. 

To have a more quantitative comparison with previous results, we apply an optical lattice $V=V_0\sin^2(x)$ and consider the tight-binding limit, as shown in Fig.~\ref{fig:bosonic}(c).
In the lowest Wannier band, $\hat\psi_{\pm 1}(x)=\sum_n W(x-n\pi)\hat a_{\pm 1,n}$. 
Choosing real $\Phi_0$ and $W$ and keeping up to the nearest neighbor hopping and on-site pairing terms, the effective Hamiltonian takes the form of Eq.~\eqref{eq:ladder}, with $J=h_1+(c_0+c_2)M_1$, $\epsilon_0=h_0+(c_0+c_2)M_0+q-\bar\mu$, and $\Delta_0=c_2M_0$, where (supplementary materials)
\begin{equation}
    \begin{split}
        h_n&=\int W(x)[-\partial_x^2+V]W(x-n\pi)dx,\\ M_n&=\int \Phi_0^2(x)W(x)W(x-n\pi)dx.
    \end{split}
\end{equation}
After preparing the polar ground state, we apply a linear potential $(\omega/\pi)\int x(|\hat \psi_1|^2-|\hat \psi_{-1}|^2)dx$ and perform the real-time GP simulation.
Non-compact Bloch oscillations in the $m_F=\pm 1$ modes are expected, with parametric amplification of a small excitation.
We introduce a small seed and fit the growth rate from the early-time population in $m_F=\pm 1$ modes, when the depletion in $m_F=0$ and nonlinear interactions in $m_F=\pm 1$ are negligible~\cite{Klempt2010}. 
As shown in Fig.~\ref{fig:bosonic}(d), the numerical growth rate (blue solid curve) is compared with predictions from the underlying $\mathrm{SU}(1,1)$ spin precession (green dashed curve), $\gamma=2{\rm log}(\rho)/T_B$, where $\rho$ denotes the largest absolute eigenvalue of $\hat{\mathcal{U}}$ determined according to Eqs.~(\ref{eq:order}, \ref{eq:ladder}).
The number, locations and strengths of the resonances are well captured. 
Including next-nearest-neighbor hopping and pairing and $M>1$ Wannier bands improves the accuracy, which also provides a means to realize other non-compact gauge fields (red dotted curve).

In summary, we demonstrate a geometric framework based on non-compact gauge groups that unifies a wide range of dynamical behaviors in non-Hermitian systems and quadratic bosonic systems.
In the presence of an electric field, it predicts non-Abelian, non-compact Bloch oscillations, which can be used for geometric amplification, quantum control, and probing distinct topological defects in non-Euclidean parameter spaces. 
It also provides a useful perspective for understanding entanglement dynamics in bosonic systems, and for performing entanglement and quantum control, which could be useful for continuous variable quantum computation and quantum-enhanced sensing. 
We hope our results will stimulate further studies of using geometric methods to explore open quantum systems, unify distinct dynamics and simulate non-compact gauge structures.

\acknowledgements{
This work is supported by the National Key R\&D Program of China (2022YFA1403700).
}

\onecolumngrid
\newpage
\vspace{0.4in}
\centerline{\bf\large Supplementary Materials for }
\centerline{\bf\large ``Dynamical effects induced by Synthetic $\mathrm{SU}(1,1)$ gauge fields''}
\setcounter{equation}{0}
\setcounter{figure}{0}
\setcounter{table}{0}
\makeatletter
\renewcommand{\theequation}{S\arabic{equation}}
\renewcommand{\thefigure}{S\arabic{figure}}
\renewcommand{\thetable}{S\arabic{table}}
\vspace{0.2in}

\section{Stability and the spectrum of an auxiliary potential}
In this section, we show that the stability condition of the non-Abelian non-compact Bloch oscillation is directly connected to the spectrum of an auxiliary periodic potential. 
In the main text, we show that the internal SU(1,1) spin evolves according to a time-dependent Hamiltonian 
\begin{equation}
 \hat H = \xi_0(t)\hat K_0 + \xi_1(t)\hat K_1 + \xi_2(t)\hat K_2, \quad \hat H(t)=\hat H(t+nT_B),
\end{equation}
where $T_B=2\pi/|\omega|$, and $\xi^2 = \xi_0^2-\xi_1^2-\xi_2^2$.
Using the representation $\hat K_0=-\hat \sigma_y/2$, $\hat K_1=-i\hat \sigma_z/2$ and $\hat K_2=i\hat \sigma_x/2$, differentiating the time-dependent Schr\"odinger equation gives
\begin{equation}
 \partial_t^2 \begin{pmatrix}\psi_1\\\psi_2\end{pmatrix} = -\frac{1}{4}\xi^2\begin{pmatrix}\psi_1\\\psi_2\end{pmatrix} + \frac{1}{2}\bigg[\partial_t\begin{pmatrix}
   -\xi_1 & \xi_2 + \xi_0 \\ 
   \xi_2 - \xi_0 & +\xi_1
 \end{pmatrix}\bigg]\begin{pmatrix}\psi_1\\\psi_2\end{pmatrix}.\label{eqs:SU11dyn}
\end{equation}
Substituting $\partial_t\psi_1 = \frac{1}{2}(-\xi_1\psi_1+(\xi_2+\xi_0)\psi_2)$ into the first row of Eq.~\eqref{eqs:SU11dyn} yields
\begin{equation}
 \partial_t^2\psi_1-\partial_t\ln|v|\partial_t\psi_1=\bigg[-\frac{1}{4}\xi^2-\frac{1}{2}(\partial_t-\partial_t\ln|v|)\xi_1\bigg]\psi_1,
\end{equation}
where $v=\xi_0+\xi_2$.
Zeros of $v$ can be avoided in a rotating frame.
Defining $\tilde\psi = e^{-i\Omega t\hat K_0}\psi$ gives $\tilde\xi_0=\xi_0+\Omega$, $\tilde\xi_1=\cos(\Omega t)\xi_1-\sin(\Omega t)\xi_2$ and $\tilde\xi_2=\cos(\Omega t)\xi_2+\sin(\Omega t)\xi_1$. Choosing a sufficiently large $\Omega/\omega\in\mathbb{N}$ ensures $\tilde v>0$ and leaves the stability condition unchanged, since $e^{-i\Omega T_B\hat K_0}=\pm\mathbb{I}_2$. We assume $v>0$ in the following discussion.

The first-order derivative term $(\partial_t\ln v)\partial_t\psi_1$ can be eliminated by a similar transformation $\psi_1 = \sqrt{v}\phi_1$. The equation of motion for $\phi_1$ becomes a time-independent Schr\"odinger equation at zero energy,
\begin{equation}
  -\partial_t^2\phi_1+V(t)\phi_1=0,\quad V = -\frac{1}{4}\xi^2-\frac{1}{2}(\partial_t-\partial_t\ln v)\xi_1-\frac{1}{2}\mathcal{S}\bigg[\int^t v(s) ds\bigg],\label{eqs:TISE}
\end{equation}
where $\mathcal{S}[\int^t v(s) ds] = v''/v-3/2(v'/v)^2$ denotes the Schwarzian derivative of $\int vdt$.

For $(\xi_0,\xi_1,\xi_2)=(2\mu,0,4\Delta\sin\omega t)$, a $\pi/2$-rotation $e^{-i\hat K_0\pi/2}$ transforms the coefficients into $(2\mu,-4\Delta\sin\omega t,0)$. 
Then, $v=2\mu$ and
\begin{equation}
 V_\mu(t) = 4\Delta^2\sin^2(\omega t)+2\Delta\omega\cos(\omega t)-\mu^2.
\end{equation}
This recovers the Whittaker-Hill potentials in the main text.
The dependence on $\mu$ is transparent after moving the constant shift to the right-hand side of Eq.~\eqref{eqs:TISE},
\begin{equation}
  -\partial_t^2\phi_1+V_0(t)\phi_1=\mu^2\phi_1.
\end{equation}
For fixed $\Delta$ and $\omega$, the potential $V_0$ is independent of $\mu$. 
A single calculation therefore determines the stability for each $\mu^2$ by evaluating its discriminant at $E = \mu^2$. A band interior gives bounded evolution, a band gap gives exponential instability.

\begin{figure}[h]
  \includegraphics[width=0.8\textwidth]{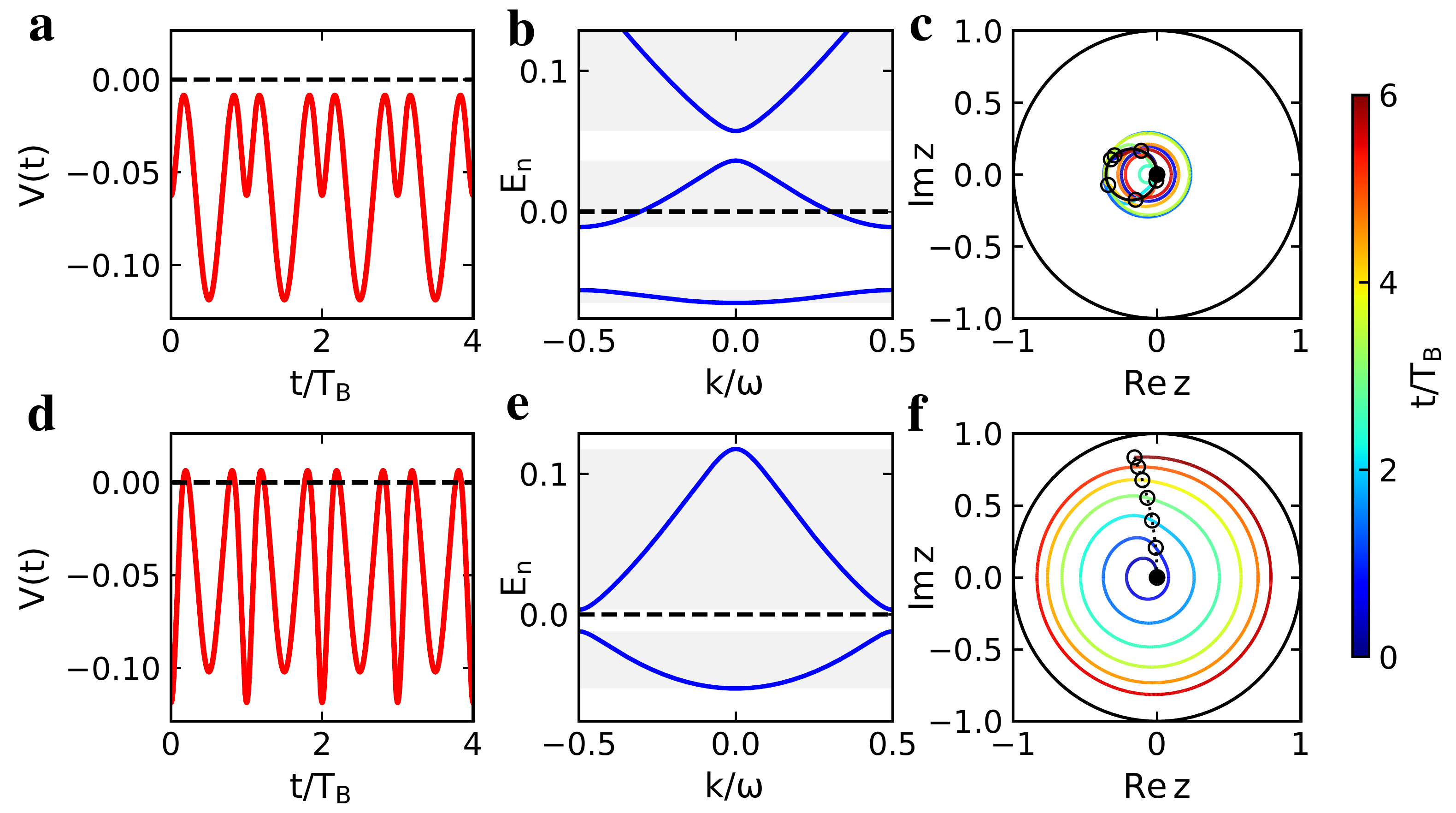}
  \caption{(a,d) Potential profiles in Eq.~\eqref{eqs:auxiliary} and (b,e) the corresponding energy spectra for $\epsilon_0=0.225$, $J=-0.0734$, $\Delta_0 = 0.0457$ with $\omega = 0.1\pi$ for (a-c) and $0.14\pi$ for (d-f). (c,f) Trajectories on the Poincar\'e disk for the Hamiltonian $\hat H =[2\epsilon_0+4J\cos(\omega t)]\hat K_0+2\Delta_0\hat K_2$ with a total evolution time $6T_B=12\pi/|\omega|$.}
  \label{Figs:potential}
\end{figure}

It is worth noting that such an auxiliary potential is not unique. Without the $\pi/2$-rotation, $v=2\mu+4\Delta\sin(\omega t)$. The auxiliary potential becomes 
\begin{equation}
 V(t) = \Delta \left(\frac{\omega ^2 \sin (\omega t)}{\mu +2 \Delta \sin (\omega t)}+\frac{3 \Delta \omega ^2 \cos ^2(\omega t)}{(\mu +2 \Delta \sin (\omega t))^2}+4 \Delta \sin ^2(\omega t)\right)-\mu ^2.
\end{equation}
When $(\xi_0,\xi_1,\xi_2)=(2\epsilon_0+4J\cos\omega t,0, 2\Delta_0)$, we have $v=2\epsilon_0+2\Delta_0 +4J\cos(\omega t)$. The auxiliary potential is 
\begin{equation}
 V(t) = \frac{3 J^2 \omega ^2 \sin ^2(\omega t)}{\left(2 J \cos (\omega t)+\Delta_0 +\epsilon_0 \right)^2}+\frac{J \omega ^2 \cos (\omega t)}{2 J \cos (\omega t)+\Delta_0+\epsilon_0 }-(2 J \cos (\omega t)+\epsilon_0 )^2+\Delta_0^2.\label{eqs:auxiliary}
\end{equation}
This corresponds to the spinor condensate under mean-field and tight-binding approximations in the main text, where we retain nearest-neighbor hopping, onsite energy, and onsite pairing terms.
Under the same parameters, we use $\epsilon_0=0.225$, $J=-0.0734$ and $\Delta_0 = 0.0457$, which are derived in the last section.
Fig.~\ref{Figs:potential} plots the potential, spectrum, and trajectory on the Poincar\'e disk at $\omega = 0.1\pi$ and $0.14\pi$, respectively, showing that stable and unstable dynamics correspond, respectively, to zero lying in a band and a gap of Eq.~\eqref{eqs:auxiliary}.

\section{Solid angle in Minkowski space}
Similar to spherical coordinates in 3-dimensional Euclidean space, the 2+1-dimensional Minkowski space can be parameterized as $\vec X_i=\xi(\pm\cosh(\eta),\sinh(\eta)\cos(\varphi),\sinh(\eta)\sin(\varphi))^T$ or $\vec X_o=\bar\xi(\sinh(\eta),\cosh(\eta)\cos(\varphi),\cosh(\eta)\sin(\varphi))^T$ inside or outside the light-cone, respectively.
The constant-$\xi$($\bar\xi$) surfaces are two(one)-sheeted hyperboloids.

Differentiating $\vec X_o$ gives 
\begin{equation}
 \begin{split}
  d\xi_0 =& \sinh(\eta)d\bar\xi + \bar\xi\cosh(\eta)d\eta,\\
  d\xi_1 =& \cosh(\eta)\cos(\varphi)d\bar\xi + \bar\xi\sinh(\eta)\cos(\varphi)d\eta - \bar\xi\cosh(\eta)\sin(\varphi)d\varphi,\\
  d\xi_2 =& \cosh(\eta)\sin(\varphi)d\bar\xi + \bar\xi\sinh(\eta)\sin(\varphi)d\eta + \bar\xi\cosh(\eta)\cos(\varphi)d\varphi.\\
 \end{split}
\end{equation}
Substitution into the Cartesian metric $ds^2=-d\xi_0^2+d\xi_1^2+d\xi_2^2$ yields the metric in the pseudo-spherical coordinates 
\begin{align}
 ds^2=&+d\bar\xi^2-\bar\xi^2d\eta^2+\bar\xi^2\cosh(\eta)^2d\varphi^2.
\end{align}
At constant $\bar\xi$, $(\eta,\varphi)$ parameterizes a 1+1 dimensional de Sitter space. Its surface element is $\sqrt{|\det(g)|}=\bar\xi^2\cosh(\eta)$, and the hyperbolic solid angle is $d\Omega_M=\cosh(\eta)d\eta d\varphi$.
In the main text, we calculated the Berry connection $A=-i\sinh(\eta)d\varphi/2$. 
Direct exterior derivative gives the Berry curvature $dA=F=-i\cosh(\eta)d\eta\wedge d\varphi/2$~\cite{Nakahara2018}. 
Therefore $\int F=(-i/2)\int d\Omega_M$. 
Consequently, the geometric contribution to the adiabatic wavefunction norm is related to the area on a constant-$\bar\xi$ surface and its hyperbolic solid angle.

Alternatively, we can transform $A$ into Cartesian coordinates, $A=A_id\xi_i$.
In components, it reads
\begin{equation}
 \vec A=\frac{i\xi_0}{2(\xi_1^2+\xi_2^2)\sqrt{\xi_1^2+\xi_2^2-\xi_0^2}}\bigg(0,\xi_2,-\xi_1\bigg)^T.
\end{equation}
Taking the curl of $\vec A$, we obtain
\begin{equation}
 \nabla\cross \vec A= \frac{i}{2(\xi_1^2+\xi_2^2-\xi_0^2)^{3/2}}(\xi_0,\xi_1,\xi_2)^T,
\end{equation} 
which satisfies $\int\nabla\cross \vec A\cdot(\partial_\eta \vec X_o\cross\partial_\varphi\vec X_o) d\eta d\varphi=(-i/2)\int d\Omega_M$ and is consistent with the previous result.

For the trajectory considered in the main text, the loop at constant $\eta$ and height $\xi_0=2\mu$ is compared with the reference loop at $\eta=0$. Here $\bar\xi=2\Gamma$.
The annulus between them gives
\begin{equation}
  \int F = (-i/2)2\pi\int_0^{\eta}d\eta'\cosh(\eta') = -i\pi\sinh(\eta),
\end{equation}
where $\sinh(\eta) = \mu/\Gamma$. The geometric amplification of the norm of the wavepacket after one cycle is $e^{2\pi\mu/\Gamma}$.

\section{SU(1,1) gauge transformation at $\Delta>J$}
For a finite open chain with $\Delta>J\geq 0$, the bosonic Kitaev chain cannot be transformed into a tight-binding model through unitary transformations. 
Instead, we can define a unitary transformation such that the resulting Hamiltonian becomes a purely squeezing model
\begin{equation}
 \hat H_{b,{\rm eff}} = \Delta_{\rm eff}\sum_{n=0}^{L-2}(\hat a^\dag_n\hat a^\dag_{n+1}+h.c.),
\end{equation}
where $\Delta_{\rm eff}=\sqrt{\Delta^2-J^2}$.
To construct the transformation, we use the Nambu-Bogoliubov spinor $\hat \psi_n=(\hat a_n,\hat a_n^\dag)^T$ and
\begin{equation}
  i\partial_t\hat \psi_n = \hat M_-\hat\psi_{n-1}+\hat M_+\hat\psi_{n+1},\quad \hat M_\pm=\pm iJ\sin(\theta)\mathbb{I}_2+\hat M
\end{equation}
where $\hat M = 2(J\cos(\theta)\hat K_0+\Delta(\sin(\phi)\hat K_1+\cos(\phi)\hat K_2))$, $\hat K_0=\hat \sigma_z/2,\hat K_1=i\hat \sigma_x/2,\hat K_2=i\hat \sigma_y/2$.
Define the two auxiliary transformations,
\begin{equation}
  \hat U_1 = e^{+i\hat K_0\phi},\quad \hat U_2 = e^{+i\hat K_1\eta_1},\quad \eta_1=\frac{1}{2}\ln(\frac{\Delta +J\cos(\theta)}{\Delta-J\cos(\theta)}),
\end{equation}
such that 
\begin{equation}
 \begin{split}
  \hat U_2^{-1}\hat U_1^{-1}\hat M_\pm \hat U_1 \hat U_2=e^{\mp i\eta_2 \hat K_2}(2\sqrt{\Delta^2-J^2}\hat K_2),\quad \eta_2=2{\rm arcsinh}\bigg(\frac{J\sin(\theta)}{\sqrt{\Delta^2-J^2}}\bigg).
 \end{split}
\end{equation}
Therefore, a site-dependent unitary transformation which is unstable in $|n|$,
\begin{equation}
  \hat{\mathcal{G}}_n = e^{-in\eta_2 \hat{\mathcal K}_1}e^{+i\hat{\mathcal K}_2\eta_1}e^{-i\hat{\mathcal K}_0\phi},
\end{equation} 
with $\hat{\mathcal{K}}_i$ given in the main text, leads to 
\begin{equation}
 \hat{\mathcal{G}}_n\hat \psi_n\hat{\mathcal{G}}_n^\dag = \hat U_1\hat U_2e^{+in\eta_2 \hat K_2}\hat \psi_n =\bigg(\frac{\hat M_-}{i\sqrt{\Delta^2-J^2}}\bigg)^n \hat U_1 \hat U_2\hat \sigma_y^n\hat \psi_n.
\end{equation}
It transforms the original equation of motion into
\begin{equation}
  \begin{split}
   i\partial_t\hat \psi_n = 2\sqrt{\Delta^2-J^2}\hat K_2(\hat\psi_{n-1}+\hat\psi_{n+1}).\\
  \end{split}
\end{equation}
Comparing with the original equation of motion, we find the transformed parameters are $J_{\rm eff}=0$, $\phi_{\rm eff}=0$ and $\Delta_{\rm eff}=\sqrt{\Delta^2-J^2}$.

\section{Entanglement entropy of a single delocalized supermode}
Consider a normalized collective supermode 
\begin{equation}
 \hat b=\sum_n b_n \hat a_n,\quad \sum_n|b_n|^2=1,\quad \hat{\mathcal{S}}_b(r) = e^{r(\hat b^{\dag 2}-\hat b^2)/2}.
\end{equation}
For a bipartition $A\cup \bar A$, we split $\hat b$ into two orthonormal modes,
\begin{equation}
 \hat b = \sqrt{\eta}\hat b_A+\sqrt{1-\eta}\hat b_{\bar A},\quad \hat b_A=\frac{1}{\sqrt{\eta}}\sum_{n\in A} b_n \hat a_n,\quad \hat b_{\bar A}=\frac{1}{\sqrt{1-\eta}}\sum_{n\in \bar A} b_n \hat a_n,
\end{equation}
where $\eta=\sum_{n\in A} |b_n|^2$. 
The squeezing operator can be effectively written as 
\begin{equation}
 \hat{\mathcal{S}}_b=\hat{\mathcal{R}}e^{r(\hat b_A^{\dag 2}-\hat b_A^2)/2}\hat{\mathcal{R}}^\dag,\quad \hat{\mathcal{R}}=e^{(\hat b_A \hat b_{\bar A}^\dag -\hat b_{\bar A}\hat b_A^\dag)\arccos(\sqrt{\eta})}.
\end{equation} 
If the initial state is the multi-mode vacuum, $\hat{\mathcal{S}}_b$ is equivalent to sending a single-mode squeezed state of $\hat b_A$ through a beam splitter mixing $\hat b_A$ and $\hat b_{\bar A}$ of transmissivity $\eta$. 
All other modes remain in vacuum.
Instead of working with states, we compute the symmetrized covariance matrix of the $\hat b_A$ mode~\cite{Adesso2007,Lee2024S}
\begin{equation}
 C_A = \begin{pmatrix}
  \langle \hat q_A^2\rangle & \langle \{\hat q_A,\hat p_A\}\rangle/2\\
  \langle \{\hat p_A, \hat q_A\}\rangle/2 & \langle \hat p_A^2\rangle
 \end{pmatrix}=\frac{1}{2}\begin{pmatrix}
  \eta e^{+2r}+(1-\eta) & 0 \\ 0 & \eta e^{-2r}+(1-\eta)
 \end{pmatrix},
\end{equation}
where $\hat q_A=(\hat b_A+\hat b^\dag_A)/\sqrt{2}$, $\hat p_A=(\hat b_A-\hat b^\dag_A)/(\sqrt{2}i)$, and the expectation value is taken in $\hat{\mathcal{S}}_b|{\rm vacuum}\rangle$. 

After an additional local squeezing of mode $\hat b_A$, $C_A$ becomes proportional to identity. The reduced state on $A$ is, therefore, a squeezed thermal state. 
The bipartite entanglement entropy, which is related to the thermal occupation number $n_{\rm th}=\sqrt{\langle \hat q_A^2\rangle \langle \hat p_A^2\rangle}-1/2$, is written as 
\begin{equation}
 S=(n_{\rm th}+1)\ln(n_{\rm th}+1)-n_{\rm th}\ln(n_{\rm th}).
\end{equation}
When $\eta=1/2$, $n_{\rm th}=\sinh^2(r/2)$, and the entanglement entropy is 
\begin{equation}
 S(\eta=1/2)=\ln(\sinh^2(r/2)+1)+\sinh^2(r/2)\ln(\frac{\sinh^2(r/2)+1}{\sinh^2(r/2)}),
\end{equation}
which approaches $r-\log(4/e)$ when $r\to\infty$.

\section{Higher Wannier bands}

Including higher Wannier bands and next-nearest-neighbor coupling yields better agreement between the observed parametric amplification and the growth rate predicted by group-theoretic methods. 
This section provides the details of the derivation. 

For a periodic potential with period $\pi$, including $M$ Wannier bands and using the mean-field approximation, the field operators should be expanded as 
\begin{equation}
 \hat\psi_{\pm 1} = \sum_{n,m}W^{(m)}(x-n\pi)\hat a_{\pm 1,n}^{(m)},
\end{equation}
$\hat\psi_0$ is replaced by $\Phi_0(x)e^{-i\bar\mu t}$, where $\Phi_0$ is the Gross-Pitaevskii (GP) ground state wavefunction, and $\bar\mu$ is the chemical potential. $\Phi_0$ and $\bar\mu$ are obtained by solving the GP equation
\begin{equation}
  \bar\mu\Phi_0 = (-\partial_x^2+V+c_0|\Phi_0|^2)\Phi_0.
\end{equation}
The Wannier functions satisfy
\begin{equation}
 \int dx W^{(m)}(x-j\pi)W^{(n)}(x-k\pi)=\delta_{mn}\delta_{jk}.
\end{equation}
$W^{(m)}(x)$ and $\Phi_0(x)$ are assumed to be real. 

The spin-1 Bose-gas Hamiltonian in the main text~\cite{Ho1998S, Kawaguchi2012S} can be rewritten as
\begin{equation}
 \begin{split}
  \hat H=\int dx \bigg[&\sum_{a=\pm 1,0}(\partial_x\hat \psi^\dag_a\partial_x\hat \psi_a+V\hat \psi_a^\dag\hat \psi_a)+c_0\hat \psi^{\dag}_{0}\hat \psi^{\dag}_{0}\hat \psi_{0}\hat \psi_{0}/2 +q(\hat \psi_1^\dag\hat \psi_1+\hat \psi_{-1}^\dag\hat \psi_{-1})\\+&(c_0+c_2)(
  \hat \psi^{\dag}_{+1}\hat \psi_{+1}\hat \psi^{\dag}_{0}\hat \psi_{0}+
  \hat \psi^{\dag}_{-1}\hat \psi_{-1}\hat \psi^{\dag}_{0}\hat \psi_{0})+c_2
  (\hat\psi^{\dag}_{0}\hat \psi^{\dag}_{0}\hat \psi_{+1}\hat \psi_{-1}+
  \hat \psi^{\dag}_{+1}\hat \psi^{\dag}_{-1}\hat \psi_{0}\hat \psi_{0})
  \\+&\frac{c_0+c_2}{2}(
  \hat \psi^{\dag}_{+1}\hat \psi^{\dag}_{+1}\hat \psi_{+1}\hat \psi_{+1}+
  \hat \psi^{\dag}_{-1}\hat \psi^{\dag}_{-1}\hat \psi_{-1}\hat \psi_{-1})+(c_0-c_2)
  \hat \psi^{\dag}_{+1}\hat \psi_{+1}\hat \psi^{\dag}_{-1}\hat \psi_{-1}\bigg].
 \end{split}
\end{equation}
Substituting $\hat\psi_{\pm 1}$ and $\hat\psi_{0}$ into the equation above and neglecting terms quartic in $\hat\psi_{\pm 1}$ gives the effective Hamiltonian of the initially unoccupied $m_F=\pm 1$ modes, 
\begin{equation}
 \begin{split}
  \hat H_{\rm eff} =\sum_{j,k,m,n}& [h_{k,mn}+(c_0+c_2)M_{k,mn}+(q-\bar\mu)\delta_{mn}\delta_{k0}]\big(\hat a_{+1,j+k}^{(m)\dag} \hat a_{+1,j}^{(n)} + \hat a_{-1,j+k}^{(m)\dag} \hat a_{-1,j}^{(n)}\big)\\
  +&c_2M_{k,mn}\big(\hat a_{+1,j+k}^{(m)\dag} \hat a_{-1,j}^{(n)\dag}+\hat a_{+1,j+k}^{(m)} \hat a_{-1,j}^{(n)}\big)
 \end{split}
\end{equation}
where the oscillating term $e^{-2i\bar\mu t}$ is eliminated by changing into a rotating frame, and 
\begin{equation}
 \begin{split}
  h_{k,mn} =& \int dx W^{(m)}(x)[-\partial_x^2+V] W^{(n)}(x+k\pi),\\
  M_{k,mn} =& \int dx W^{(m)}(x)\Phi_0^2(x) W^{(n)}(x+k\pi).
 \end{split}
\end{equation}
Similarly, the linear Zeeman gradient also leads to inter-band mixing, with the matrix elements 
\begin{equation}
  A_{k,mn} = \int dx xW^{(m)}(x)W^{(n)}(x+k\pi).\\
\end{equation}
The opposite linear potentials on $m_F=\pm 1$ project to 
\begin{equation}
  \hat H_{\rm E, eff}  =\frac{\omega}{\pi}\sum_{j,k,m,n} A_{k,mn}\big(\hat a_{+1,j+k}^{(m)\dag} \hat a_{+1,j}^{(n)} - \hat a_{-1,j+k}^{(m)\dag} \hat a_{-1,j}^{(n)}\big) + \omega\sum_{j,n}j\big(\hat a_{+1,j}^{(n)\dag} \hat a_{+1,j}^{(n)}-\hat a_{-1,j}^{(n)\dag} \hat a_{-1,j}^{(n)}\big).
\end{equation}

The Wannier functions $W^{(m)}(x)$ are constructed by solving the Bloch problem,
\begin{equation}
 [-\partial_x^2+V_0\sin^2(x)]w^{(m)}_\kappa(x)=E_\kappa^{(m)}w^{(m)}_\kappa(x),
\end{equation}
where $w^{(m)}_\kappa(x+\pi)=e^{+i\kappa\pi}w^{(m)}_\kappa(x)$ is the Bloch wavefunction with quasi-momentum $\kappa$ in the $m$-th band. $W^{(m)}(x) = \int_{-1}^{+1} w^{(m)}_\kappa(x)d\kappa$, up to normalization and the phase fixing of $w^{(m)}_\kappa$~\cite{David2012}.
Here, we solve $w^{(m)}_\kappa(x)$ in the momentum space, $w^{(m)}_\kappa(x)=\sum_{l=-l_{\rm max}}^{+l_{\rm max}}c^{(m)}_{\kappa,l} e^{i(\kappa+2l)x}$, such that
\begin{equation}
 \bigg[(\kappa+2l)^2+\frac{V_0}{2}\bigg]c^{(m)}_{\kappa,l}-\frac{V_0}{4}\big(c^{(m)}_{\kappa,l-1}+c^{(m)}_{\kappa,l+1}\big)=E_\kappa^{(m)}c^{(m)}_{\kappa,l}
\end{equation}
where $l_{\rm max}$ is the momentum cut-off. 
The phase is fixed by letting $c^{(m)}_{\kappa,-l_{\rm max}}$ be real and positive.

\begin{figure}[h]
  \includegraphics[width=0.8\textwidth]{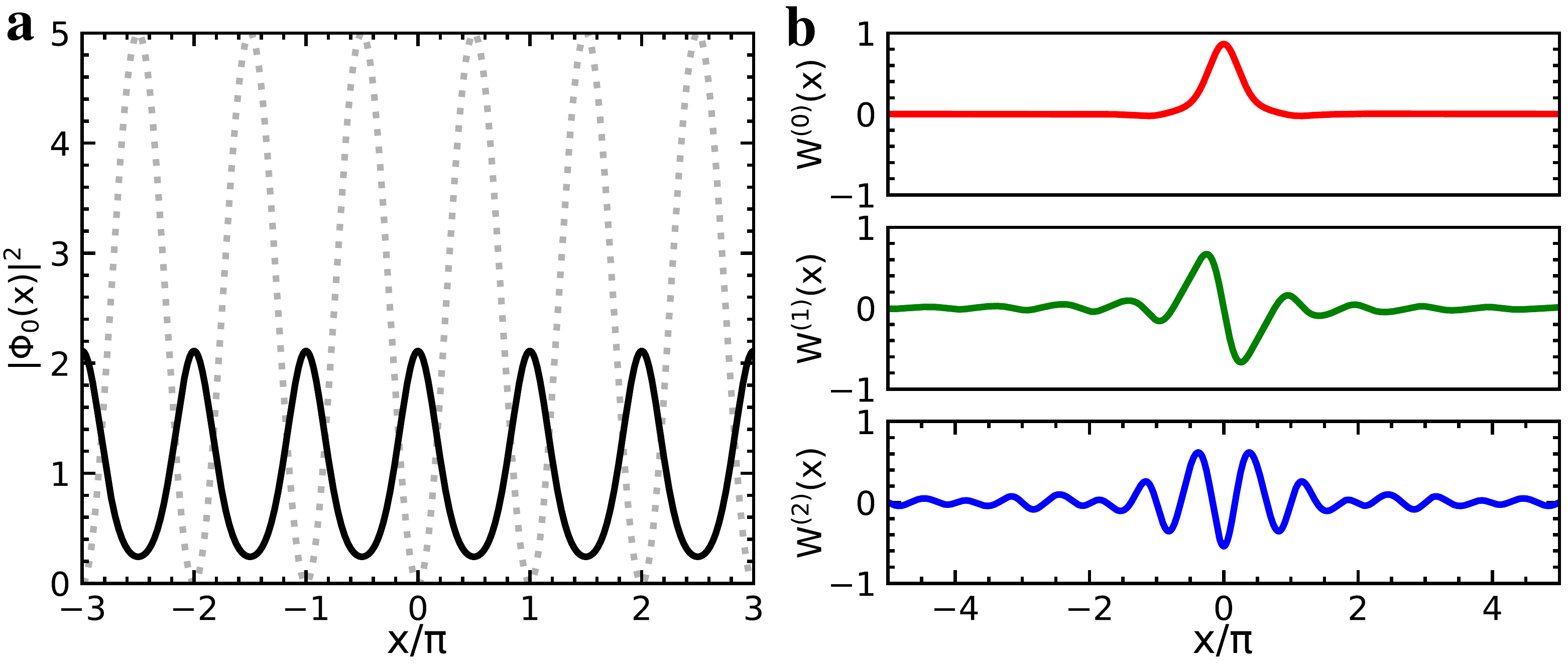}
  \caption{(a) Condensate ground-state density $|\Phi_0(x)|^2$ in $m_F=0$ without the additional linear potentials. 
   The mean density is set to unity. 
   The gray dashed curve represents the background potential. 
   (b) From top to bottom: the first three Wannier wavefunctions at $V=5\sin^2(x)$. }
  \label{Figs:Wannier}
\end{figure}

In the main text, we set $V=5\sin^2(x)$, such that multiple resonances are visible within the tuned $\omega$ parameter region. 
For this potential, with $c_0=0.2$, $c_2=0.03$, and unit mean density $\int_0^\pi|\Phi_0(x)|^2dx=\pi$, Fig.~\ref{Figs:Wannier} shows the GP ground state density $|\Phi_0(x)|^2$ obtained by imaginary time evolution and the lowest three Wannier functions. Numerically, $\bar\mu=2.109$.
With $q=0.04$, keeping only nearest-neighbor hopping, onsite energy, and onsite pairing terms, the parameters $\epsilon_0=0.225$, $J=-0.0734$ and $\Delta_0 = 0.0457$ in the first section are obtained. 
For up to three Wannier bands and next-nearest neighbor couplings, the calculated values of $h_{k,mn}$, $M_{k,mn}$ and $A_{k,mn}$ are listed as below.  $X_{-k}=X_{+k}^T$ for $X=h,M,A$.
\begin{equation}
  \begin{split}
    h_0 =& \begin{pmatrix}
   1.944 &0 &0\\
 0 &5.264 &0\\
 0 &0 &9.021\\
  \end{pmatrix},\quad 
  h_{-1} = \begin{pmatrix}
    -0.0658&0 &0\\
  0 & 0.423  &0\\
  0& 0 &-0.958\\
  \end{pmatrix},\\
  h_{-2} =& \begin{pmatrix}
    0.00340 &0  &0\\
  0  &0.0728 &0\\
  0 & 0  &0.0736\\
  \end{pmatrix}.
  \end{split}
\end{equation}

\begin{equation}
  \begin{split}
    M_0 =& \begin{pmatrix}
    1.522 & 0 & -0.433 \\ 
    0 & 0.807 & 0 \\ 
    -0.433 & 0 & 0.870 \\
  \end{pmatrix},\quad 
  M_{-1} = \begin{pmatrix}
    -0.0332  &-0.144 &-0.0298\\
    0.144  &0.0637 &0.124\\
    -0.0298  &-0.124 &-0.0355\\
  \end{pmatrix},\\
  M_{-2} =& \begin{pmatrix}
   0.00385&  -0.0337&  -0.00628\\
 0.0337&   0.0103&   -0.0293\\
 -0.00628& 0.0293& -0.0126\\
  \end{pmatrix}.
  \end{split}
\end{equation}

\begin{equation}
  \begin{split}
    A_0 =& \begin{pmatrix}
    0 &-0.544 & 0\\
 -0.544&  0 &-0.773\\
  0& -0.773 &0\\
  \end{pmatrix},\quad 
  A_{-1} = \begin{pmatrix}
    0 & 0.135 & -0.0487\\
  0.135&  0 &-0.594\\
 0.0487& -0.594 &0\\
  \end{pmatrix},\\
  A_{-2} =& \begin{pmatrix}
   0 &-0.0191  &-0.0135\\
 -0.0191 & 0& -0.444\\
 0.0135& -0.444&  0\\
  \end{pmatrix}.
  \end{split}
\end{equation}

\end{document}